\documentclass[preprint]{emulateapj} 

\usepackage{natbib}
\usepackage{hyperref}
\hypersetup{colorlinks,citecolor=Blue,linkcolor=Red,urlcolor=Blue}
\usepackage{amsmath}
\usepackage{empheq}
\usepackage{mathrsfs}
\usepackage[usenames,dvipsnames]{color}

\newcommand{\Bo}{\vec{B}_0} 
\newcommand{\Be}{\vec{B}_{\epsilon}} 
\newcommand{\vo}{\vec{v}_0} 
\newcommand{\ve}{\vec{v}_{\epsilon}}

\shorttitle{Magnetically Perturbed Jets} 

\begin{document}

\title{Non-Axisymmetric Flows on Hot Jupiters with Oblique Magnetic Fields}  
\author{Konstantin Batygin$^1$ and Sabine Stanley$^2$} 

\affil{$^1$Division of Geological and Planetary Sciences, California Institute of Technology, 1200 E. California Blvd., Pasadena, CA 91125}
\affil{$^2$Department of Physics, University of Toronto, 60 St. George St., Toronto, ON} 


\email{kbatygin@gps.caltech.edu}

\begin{abstract}

Giant planets that reside in close proximity to their host stars are subject to extreme irradiation, which gives rise to thermal ionization of trace Alkali metals in their atmospheres. On objects where the atmospheric electrical conductivity is substantial, the global circulation couples to the background magnetic field, inducing supplementary fields and altering the nature of the flow. To date, a number of authors have considered the influence of a spin-pole aligned dipole magnetic field on the dynamical state of a weakly-ionized atmosphere and found that magnetic breaking may lead to significantly slower winds than predicted within a purely hydrodynamical framework. Here, we consider the effect of a tilted dipole magnetic field on the circulation and demonstrate that in addition to regulating wind velocities, an oblique field generates stationary non-axisymmetric structures that adhere to the geometry of the magnetic pole. Using a kinematic perturbative approach, we derive a closed-form solution for the perturbed circulation and show that the fractional distortion of zonal jets scales as the product of the field obliquity and the Elsasser number. The results obtained herein suggest that on planets with oblique magnetic fields, advective shifts of dayside hotspots may have substantial latitudinal components. This prediction may be tested observationally using the eclipse mapping technique. 

\end{abstract}

\section{Introduction}

Since the discovery of hot Jupiters \citep{1995Natur.378..355M} and their detection in transit \citep{2000ApJ...529L..45C}, the study of atmospheric dynamics on these objects has attracted substantial interest. Accordingly, over the last decade a vast hierarchy of global circulation models (``GCMs", largely adopted from numerical codes aimed at simulating the Earth's climate) has been established (see \citealt{2011exop.book..471S} and the references therein). Although distinct models exhibit subtle differences (see \citealt{2011MNRAS.tmp..370H}), hydrodynamical simulations show broad agreement on the qualitative features of the flow. Specifically, the overwhelming majority of 3D GCMs obtain super-rotating eastward zonal jets with characteristic maximal speeds of order $|v|_{\rm{max}} \sim$ few km/s at photospheric (and somewhat higher) pressures.

Because atmospheric temperatures on hot Jupiters can reach values high enough to ionize alkali metals such as K and Na \citep{2010ApJ...714L.238B,2012ApJ...748L..17H}, it has been realized that magnetohydrodynamic (MHD) effects may carry substantial repercussions for global circulation. To simulate the coupling between the atmospheric flow and the background field, a number of authors have augmented their GCMs with a Rayleigh drag aimed at mimicking the coupling between the large-scale circulation and the background field \citep{2010ApJ...719.1421P, 2010ApJ...713.1174M, 2013ApJ...764..103R}. Although a natural first step, this approach is not adequate generally since magnetic torques inherently depend on the geometry of the field and exhibit phase signatures of damping that differ from those of a Rayleigh drag \citep{2014arXiv1401.7571H}. Accordingly, the first self-consistent Boussinesq magnetohydrodynamic simulations were performed by \citet{2013ApJ...776...53B} and recently, improved anaelastic MHD calculations were presented by \citet{2014arXiv1401.5815R}. By and large, this aggregate of magnetic simulations suggests that zonal jets are damped by Lorentz forces, in agreement with analytical studies \citep{2012ApJ...745..138M}.

A simplifying assumption that is consistently employed when studying MHD effects in hot atmospheres is the alignment between the magnetic axis and the spin axis of the planet. Although this is synonymous to the case of Saturn \citep{2010GeoRL..37.5201S}, other giant planets in the Solar System posses oblique magnetic fields \citep{2003E&PSL.208....1S}, and it is reasonable to expect that misalignments between the magnetic and rotational axes will be common to the population of close-in planets. Correspondingly, in this paper we address the effects of a tilted dipole magnetic field on global circulation in a partially ionized atmosphere. The paper is organized as follows. In section 2, we describe the considered physical setup. In section 3, we calculate the currents as well as the associated Lorentz forces, induced by the interactions between the background field and the imposed flow. In section 4, we examine the perturbations to the zonal jets that arise from the Lorentz forces. We conclude and discuss our results in section 5. Throughout the paper, exclusively analytical perturbative techniques are employed. 

\section{Framework}

There exists a clear trade-off between model simplicity and realism in the study of dynamical meteorology. Accordingly, while computationally intensive 3D GCMs have been utilized for constructing the most detailed representations of the atmospheric states of hot Jupiters \citep{2005ApJ...629L..45C, 2008ApJ...685.1324S, 2009ApJ...699..564S, 2009ApJ...700..887M, 2012ApJ...750...96R, 2011MNRAS.tmp..370H, 2008ApJ...673..513D, 2012arXiv1211.1709D, 2014arXiv1401.5815R}, simplified 2D simulations 
\citep{2003ApJ...587L.117C, 2008ApJ...675..817C,2007ApJ...657L.113L, 2008ApJ...674.1106L, 2011ApJ...738...71S, 2013ApJ...776...53B} have been used to elucidate and analyze specific phenomena. As this study finds itself in the latter category, we shall also restrict our treatment to a thin shell (envisioned to reside at the photospheric pressure level) of thickness 
\begin{equation}
\delta \sim H \equiv k_B T/\mu g \ll R_{\rm{p}},
\end{equation}
where $k_B$ is Boltzmann's constant, $T$ is temperature, $\mu$ is mean molecular weight, $g$ is surface gravity, and $R_{\rm{p}}$ is the planetary radius\footnote{In other words, the philosophy of the paper can be summarized as follows: we envision reality, construct a toy model that approximates reality, chop off the limbs of the toy and subsequently polish it up to a sphere \citep{Mattrchat}.}. 

\begin{figure}
\includegraphics[width=0.85\columnwidth]{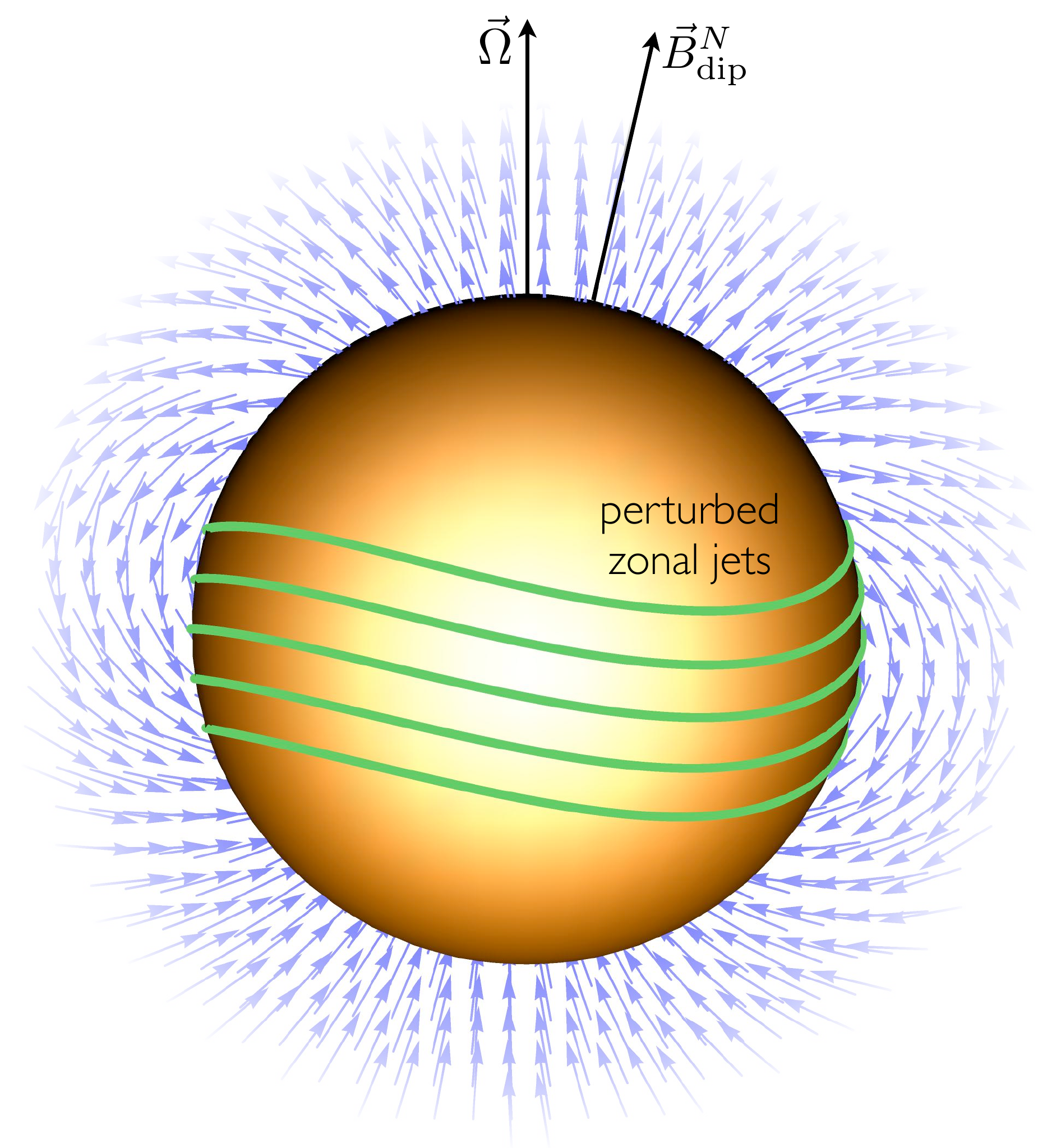}
\caption{A cartoon depicting the geometrical setup of the problem. The spin vector of the planet (labeled ${\vec{\Omega}}$) is taken to coincide with the $\hat{z}$-axis, while the background magnetic field (shown as a blue vector field) is tilted towards the $\hat{x}$-axis by a small angle, $\psi$. As a consequence of the atmosphere's finite conductivity, zonal jets (shown as green hoops) are envisioned to be perturbed away from their default azimuthally symmetric state.} 
\label{toy}
\end{figure}

For tractability, we assume that the inclination of the background magnetic field relative to the spin-axis, $\psi$, is small. Thus, we introduce:
\begin{equation}
\epsilon=\sin(\psi)\ll1
\end{equation}
as an inherent small parameter of the problem. The background dipole field then takes the form \citep{1998clel.book.....J}:
\begin{eqnarray}
\vec{B}_{\rm{dip}}&=&\Bo+\epsilon\Be=B_{\rm{p}}\vec{\nabla}\bigg[R_{\rm{p}}\left(\frac{R_{\rm{p}}}{r}\right)^2\mathcal{P}_1^0(\cos(\theta))\nonumber\\
&+&\epsilon R_{\rm{p}}\left(\frac{R_{\rm{p}}}{r}\right)^2\cos(\phi)\mathcal{P}_1^1(\cos(\theta))\bigg],
\end{eqnarray}
where $B_{\rm{p}}$ is the surface field strength, $\mathcal{P}$ is the associated Legendre polynomial and the $\hat{x}$-axis of the coordinate system is chosen to correspond to the field's line of nodes. Meanwhile, following \citet{2008Icar..196..653L, 2010ApJ...714L.238B} we adopt a kinematic prescription for the zonal jet:
\begin{eqnarray}
\vec{v}&=&\vo+\epsilon\ve=v_{\rm{max}}\sin(\theta)\hat{\phi}+\epsilon\ve,
\end{eqnarray}
where $\ve$ is an unknown perturbation to the background flow. We assume that radial flow is prohibited in the region of interest, and that $\ve$ is independent of $r$ (i.e. $\ve$ represents an azimuthally asymmetric 2D flow). Additionally, we assume that the magnetic diffusivity, $\eta$, is constant at the pressure-level of interest. A cartoon depicting the geometrical setup of the problem is shown in Figure (\ref{toy}).

\section{Magnetic Induction}

The steady state induction equation, relevant to partially ionized hot Jupiter atmospheres, is written as follows \citep{1978mfge.book.....M}:
\begin{equation}
\label{induction}
0=\eta\vec{\nabla}^2\vec{B}+\vec{\nabla}\times\left(\vec{v}\times\vec{B}\right).
\end{equation}
Note that in the above expression, the Hall and ambipolar diffusion terms have been neglected, as justified by \citet{2010ApJ...719.1421P}. The total magnetic field is composed of the sum of the background and induced fields: $\vec{B} = \vec{B}_{\rm{dip}}+\vec{B}_{\rm{ind}}$. However, since $\vec{B}_{\rm{dip}}$ is defined as a gradient of a scalar function, it is curl-free by construction meaning 
\begin{equation}
\vec{\nabla}^2\vec{B}=\vec{\nabla}^2\vec{B}_{\rm{ind}}. 
\end{equation}
On the other hand, for clarity we may take the $(\vec{v} \times \vec{B})$ term to be dominated by $\vec{B}_{\rm{dip}}$ \citep{2010ApJ...714L.238B}. Conventionally, this assumption implies a magnetic Reynolds number 
\begin{equation}
\mathcal{R}_{\rm{m}}\equiv\frac{\mathcal{V} \delta}{\eta}
\end{equation}
somewhat smaller than unity. As discussed by \citet{2012ApJ...745..138M}, magnetic Reynolds numbers of order unity or less are expected for a sizable fraction of hot Jupiters, especially under the assumptions of strong (e.g. $\sim 30$ G) magnetic fields (see also \citealt{2009Natur.457..167C}). However, it should also be noted that the aforementioned assumption may be justified even in the regime where $\mathcal{R}_{\rm{m}}$ is substantial, provided a circulation geometry that is not complex enough to directly influence $(\vec{v} \times \vec{B})$ (a simple example is an unperturbed kinematic zonal flow, that induces a purely toroidal field, which in turn drops out of the induction term in equation (\ref{induction}) - see e.g. \citealt{2013ApJ...776...53B}).

Consequently. to first order in $\epsilon$, equation (\ref{induction}) reads:
\begin{eqnarray}
\label{inductionFO}
0&=&\eta\vec{\nabla}^2\vec{B}_{\rm{ind}}^{(0,0)}+\vec{\nabla}\times\left(\vo\times\Bo\right)\nonumber\\ 
&+&\epsilon \bigg[\eta\vec{\nabla}^2\vec{B}_{\rm{ind}}^{(0,\epsilon)}+\vec{\nabla}\times\left(\vo\times\Be\right)\nonumber\\ 
&+&\eta\vec{\nabla}^2\vec{B}_{\rm{ind}}^{(\epsilon,0)}+\vec{\nabla}\times\left(\ve\times\Bo\right)\bigg].
\end{eqnarray}
In the above expression, $\vec{B}_{\rm{ind}}^{(0,0)}$ represents the field induced by the interaction of the background flow ($\vo$) with the the aligned component of the field ($\Bo$), $\vec{B}_{\rm{ind}}^{(0,\epsilon)}$ depicts the field induced by the interaction of the background flow ($\vo$) with the non-axisymmetric component of the field ($\Be$), and $\vec{B}_{\rm{ind}}^{(\epsilon,0)}$ designates the field arising from the coupling between the (unknown) perturbed component of the circulation ($\ve$) and the axisymmetric field ($\Bo$). 

As a first step towards solving equation (\ref{inductionFO}), we shall assume that magnetic induction associated with the unknown velocity field $\ve$ gives rise to a negligible Lorentz force and need not be addressed. As shown in the appendix, this simplification is equivalent to assuming that the Elsasser number, defined as:
\begin{equation}
\Lambda\equiv\frac{B_{\rm{p}}^2}{\mu_0\eta\rho\Omega},
\end{equation}
where $\rho$ is the atmospheric density and $\mu_0$ is the permeability of free space, is much less than unity. The first and second lines of equation (\ref{inductionFO}) can then be solved sequentially and independently. 

\subsection{Zeroeth-Order Solution}
The leading order solution to equation (\ref{inductionFO}) can be trivially obtained by separation of variables and has been discussed elsewhere (see e.g. \citealt{2013ApJ...776...53B}). We shall briefly rehash it here for completeness. The expression $\vec{\nabla} \times (\vo \times \Bo)$ possesses azimuthal symmetry (i.e. only has a $\hat{\phi}$ component and carries no $\phi-$dependence). Moreover, its angular part is an eigenfunction of the $\nabla^2$ operator. Correspondingly, we immediately find that the solution for $\vec{B}_{\rm{ind}}^{(0,0)}$ is satisfied by 
\begin{equation}
\vec{B}_{\rm{ind}}^{(0,0)}=f(r)\cos(\theta)\sin(\theta)\hat{\phi}.
\end{equation}
Thus, the first line of equation (\ref{inductionFO}) becomes a second-order ODE for $f(r)$:
\begin{equation}
\eta\big(6f(r)-r(2f'(r)+rf''(r))\big)=-\frac{2B_{\rm{p}}R_{\rm{p}}^3v_{\rm{max}}}{r^2}.
\end{equation}
Following \citet{2013ApJ...776...53B} and \citet{2014arXiv1401.5815R}, we adopt zero radial current boundary conditions at the outer $(r = R_{\rm{p}})$ and inner $(r = R_{\rm{p}}- \delta)$ edges of the shell: $(\vec{\nabla} \times \vec{B}_{\rm{ind}}^{(0,0)})_{\hat{r}} \propto f(r) = 0$ . The solution for $f(r)$ then takes the form:
\begin{eqnarray}
f(r)&=&B_{\rm{p}}R_{\rm{p}}^3v_{\rm{max}}\left(r-R_{\rm{p}}\right)\big(r^4+r^3R_{\rm{p}}+r^2R_{\rm{p}}^2\nonumber\\
&+&rR_{\rm{p}}^3-4R_{\rm{p}}^4+10R_{\rm{p}}^3\delta-10R_{\rm{p}}^2\delta^2+5R_{\rm{p}}\delta^3\nonumber\\
&-&\delta^4\big)/\big(2\eta r^3(5R_{\rm{p}}^4-10R_{\rm{p}}^3\delta+10R_{\rm{p}}^2\delta^2\nonumber\\
&-&5R_{\rm{p}}\delta^3+\delta^4 \big)
\end{eqnarray}

\begin{figure}
\includegraphics[width=1\columnwidth]{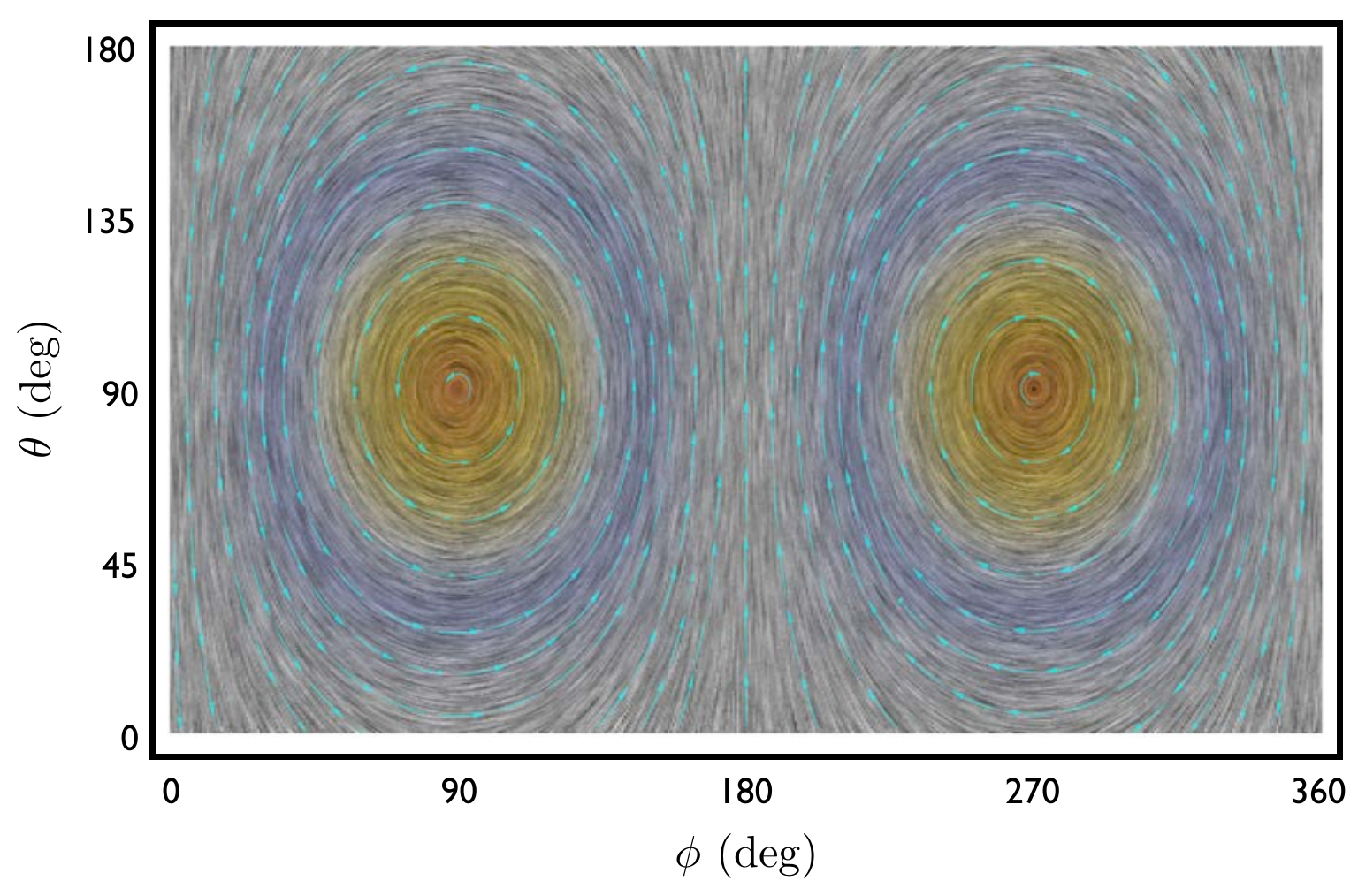}
\caption{The geometry of the vector field associated with the induced current, $\vec{J}_{\rm{ind}}^{(0,\epsilon)}$. In this figure, the radial component of the current is neglected as it is much smaller than its azimuthal and meridional counterparts. Note that the vector field's symmetry axis corresponds to the line of nodes of the tilted background magnetic field.} 
\label{current}
\end{figure}

Written out explicitly, the Lorentz force reads:
\begin{eqnarray}
\vec{F}_{\rm{L}} &=& \vec{F}_{\rm{L}\ 0}^{(0,0)} + \epsilon \vec{F}_{\rm{L} \ \epsilon}^{(0,0)} = \frac{(\vec{\nabla} \times \vec{B}_{\rm{ind}}^{(0,0)}) \times \Bo}{\rho \mu_0} \nonumber \\
&+& \epsilon \frac{(\vec{\nabla} \times \vec{B}_{\rm{ind}}^{(0,0)}) \times \Be}{\rho \mu_0}.
\label{forceref1}
\end{eqnarray}
\\
In keeping with the notation introduced above, the ubiquitous $(0,0)$ superscripts in equation (\ref{forceref1}) imply that both terms arise from currents associated with the coupling between the axisymmetric components of the flow and the field. Meanwhile, the $0$ and $\epsilon$ subscripts signify the interactions between these currents and the axisymmetric and non-axisymmetric components of the field respectively. 

Because our treatment is quasi-2D, we are primarily interested in the vertically averaged Lorentz Force:
\begin{equation}
\langle \vec{F}_{\rm{L}} \rangle = \frac{1}{\delta} \int_{R_{\rm{p}}- \delta}^{R_{\rm{p}}}\vec{F}_{\rm{L}} dr.
\end{equation}
Moreover, we shall take advantage of the smallness of the aspect ratio $\xi \equiv \delta/R_{\rm{p}} \ll 1$ and expand $\langle \vec{F}_{\rm{L}} \rangle$ as a power series. To leading order in $\xi$, the relevant expressions take the form:
\begin{eqnarray}
\label{FL}
&& \langle \vec{F}_{\rm{L}\ \epsilon}^{(0,0)} \rangle =  -\epsilon \xi^2 \frac{v_{\rm{max}} B_{\rm{p}}}{\mu_0 \eta \rho} \frac{(1+ 3 \cos(2 \theta) \sin(\phi))}{12} \hat{\theta} \nonumber \\
&& + \epsilon \xi^2 \frac{v_{\rm{max}} B_{\rm{p}}}{\mu_0 \eta \rho} \frac{(3 \cos(\theta) - 11 \cos(3\theta))\cos(\phi)}{24} \hat{\phi}.
\end{eqnarray}

Note that the $\langle \vec{F}_{\rm{L}\ 0}^{(0,0)} \rangle$ term is omitted, since it governs the conventional magnetic damping effect (see e.g. \citealt{2008Icar..196..653L,2010ApJ...719.1421P, 2011ApJ...738....1B, 2012ApJ...745..138M}) and is of no interest to us, as it is envisioned to only modulate the magnitude of $\vec{v}_{0}$ (which we take as an adjustable parameter). The term on the second line of equation (\ref{FL}) provides a small (of order $\mathcal{O}(\epsilon)$) correction to the aforementioned effect, and is also of little importance. On the contrary, the term on the first line of equation (\ref{FL}) may give rise to a qualitative alteration of the zonal jet.

\subsection{First-Order Solution}

The solution to the second line of equation (\ref{inductionFO}) is not as trivial as that considered above because the associated induction term lacks azimuthal symmetry. As such, it is difficult to obtain an analytical solution simply by inspection and we shall address this equation in a somewhat alternative manner. That is, rather than solving for the induced field $\vec{B}_{\rm{ind}}^{(0,\epsilon)}$, we shall directly seek a solution for the induced current $\vec{J}_{\rm{ind}}^{(0,\epsilon)}$. Specifically, we uncurl the second line of equation (\ref{inductionFO}) to recover Ohm's law:
\begin{equation}
\vec{J}_{\rm{ind}}^{(0,\epsilon)} = \frac{1}{\mu_0 \eta} \left( \vo \times \Be + \vec{\nabla} \Phi \right),
\end{equation}
where $\Phi$ is the electric potential. As before, the induced radial current is constrained to be null at $r = R_{\rm{p}} - \delta$ and $r = R_{\rm{p}}$. Consequently, we choose the angular part of $\Phi$ to correspond to that of $(\vo \times \Be)_{\hat{r}}$ (since it is not affected by differentiation with respect to $r$) and retain an unspecified radial dependence:
\begin{equation}
\Phi = u(r) \cos(\theta) \sin(\theta) \cos(\phi).
\end{equation}

\begin{figure*}
\includegraphics[width=1\textwidth]{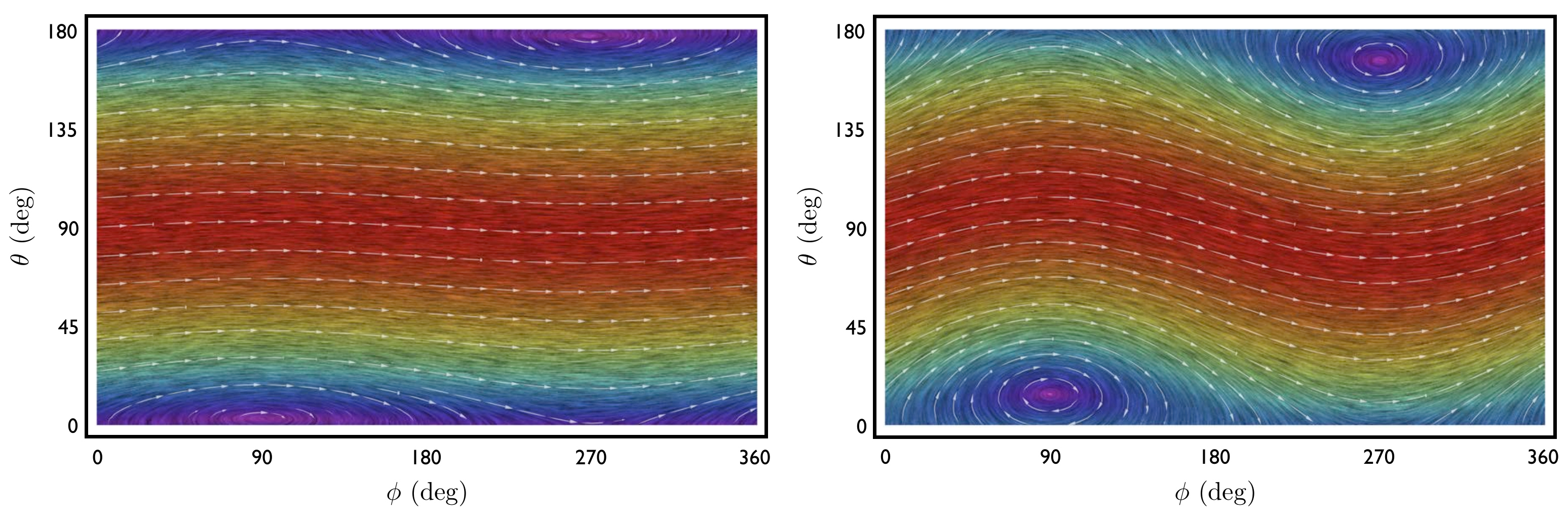}
\caption{The geometry of the perturbed zonal jet $\vo+\epsilon (\vec{\nabla} \times \Psi \hat{r})$. The left panel shows the circulation with $\epsilon \Lambda = 0.05$ and the right panel depicts the solution with $\epsilon \Lambda = 0.25$. It is expected that for systems where the deviation away from a purely zonal circulation is substantial, the dayside hotspot will be advected both east-wards and pole-wards. Naturally, the extent of the hotspot's displacement is dependent on both, the geometry of the field and the magnitude of the perturbation.} 
\label{flow}
\end{figure*}

Taking advantage of the divergence-free nature of the induced current 
\begin{equation}
\vec{\nabla} \cdot \vec{J}_{\rm{ind}}^{(0,\epsilon)} = 0,
\end{equation}
we obtain a second-order ODE for $u(r)$:
\begin{equation}
r^2 u''(r) + 2 r u'(r) - 6 u(r) = \frac{5 \epsilon B_{\rm{p}} R_{\rm{p}}^3 v_{\rm{max}}}{r^2}.
\end{equation}
With the aforementioned boundary conditions (i.e. $R_{\rm{p}}^3 u'(R_{\rm{p}})  = \epsilon v_{\rm{max}} B_{\rm{p}} R_{\rm{p}}^3$ \& $(R_{\rm{p}} - \delta)^3 u'(R_{\rm{p}} - \delta)  = \epsilon v_{\rm{max}} B_{\rm{p}} R_{\rm{p}}^3$), we obtain:
\begin{eqnarray}
u(r) &=& \epsilon B_{\rm{p}} R_{\rm{p}}^3 v_{\rm{max}} \big(2 R_{\rm{p}}( 4 R_{\rm{p}}^4 - 10 \delta R_{\rm{p}}^3 + 10 \delta^2 R_{\rm{p}}^2 \nonumber \\
 &-& 5 \delta^3 R_{\rm{p}} + \delta^4) - 5r (5 R_{\rm{p}}^4 - 10 \delta R_{\rm{p}}^3+ \delta^2 R_{\rm{p}}^2 \nonumber \\
 &-& 5 \delta^3 R_{\rm{p}} + \delta^4)  - 3 r^5  \big)/(4 r^3 (5 R_{\rm{p}}^4 - 10 \delta R_{\rm{p}}^3 \nonumber \\
 &+& 10 \delta^2 R_{\rm{p}}^2 -5 \delta^3 R_{\rm{p}}+ \delta^4)).
\end{eqnarray}
Because the thickness of the atmospheric shell in this calculation is taken to be small compared to the planetary radius, the current density $\vec{J}_{\rm{ind}}^{(0,\epsilon)}$ is primarily azimuthal and meridional. Neglecting the small radial component, the 2D geometry of $\vec{J}_{\rm{ind}}^{(0,\epsilon)}$ is shown in Figure (\ref{current}).

The corresponding Lorentz force then takes the form:
\begin{eqnarray}
\vec{F}_{\rm{L}\ 0}^{(0,\epsilon)} = \epsilon \frac{\vec{J}_{\rm{ind}}^{(0,\epsilon)} \times \Bo}{\rho}.
\end{eqnarray}
Concurrently, to leading order in $\xi$, its vertically averaged counterpart reads:
\begin{eqnarray}
\label{FLsecond}
\langle \vec{F}_{\rm{L}\ 0}^{(0,\epsilon)} \rangle &=& \epsilon \frac{2 B_{\rm{p}}^2 v_{\rm{max}}}{\mu_0 \eta \rho} \cos(\theta)^2 \sin(\phi) \hat{\theta} \nonumber \\
&+& \epsilon \frac{2 B_{\rm{p}}^2 v_{\rm{max}}}{\mu_0 \eta \rho} \cos(\theta) \sin(\phi) \hat{\phi}.
\end{eqnarray}

Cumulatively, there are two constituents of the Lorentz force, given by equations (\ref{FL}) and (\ref{FLsecond}). Their ratio is of order:
\begin{eqnarray}
\frac{\langle \vec{F}_{\rm{L}\ \epsilon}^{(0,0)} \rangle}{\langle \vec{F}_{\rm{L}\ 0}^{(0,\epsilon)} \rangle} \sim \mathcal{O}(\xi^2) \ll 1.
\end{eqnarray}
Consequently, in subsequent analysis, $\langle \vec{F}_{\rm{L}\ \epsilon}^{(0,0)} \rangle$ can be neglected in favor of $\langle \vec{F}_{\rm{L}\ 0}^{(0,\epsilon)} \rangle$. 

\section{Perturbed Circulation}

With the perturbing forces specified, we now seek a solution for the non-axisymmetric perturbation to the zonal flow. Within the context of the thin-shell geometry considered here, we can make use of the Boussinesq approximation, and neglect density variations in the domain of interest. Under this assumption, the continuity equation reduces to the incompressibility condition:
\begin{eqnarray}
\label{incompress}
\vec{\nabla} \cdot \ve = 0.
\end{eqnarray}
Owing to our 2D treatment of the global circulation, equation (\ref{incompress}) can be automatically satisfied by specifying a stream-function, $\Psi$, from which $\ve$ arises \citep{1959flme.book.....L}:
\begin{eqnarray}
\ve = \vec{\nabla} \times \Psi \hat{r}.
\end{eqnarray}

In order to obtain the correction to the velocity field, $\ve$, we must (perturbatively) solve the imvicid Navier-Stokes equation \citep{1992aitd.book.....H}:
\begin{eqnarray}
\frac{D \vec{v}}{D t} = - 2 \vec{\Omega} \times \vec{v} + \frac{\vec{\nabla} P}{\rho} + \vec{g} + \langle \vec{F}_{\rm{L}\ 0}^{(0,\epsilon)} \rangle.
\label{NavierStokes}
\end{eqnarray}
Here, $D/Dt = \partial / \partial t + \vec{v} \cdot \vec{\nabla} $ is the material derivative, $\vec{\Omega} = \Omega \hat{z}$ is the spin vector and $P$ is pressure. 

Firstly, as in the previous section, we assume steady state, meaning $\partial / \partial t \rightarrow 0$. Secondly, choosing a characteristic velocity scale of order  $\mathcal{V} \sim 1$ km/s, a characteristic length scale of order $\mathcal{L} \sim R_{\rm{p}}$ and a spin-period of order $\mathcal{T} \sim 1$ day (which translates to a Coriolis parameter of $\bar{f} = 4 \pi \cos(\pi/4) / \mathcal{T}$), we obtain a Rossby number of order
\begin{equation}
\mathcal{R}_{\rm{o}} \equiv \frac{\mathcal{V}}{\mathcal{L} \bar{f}} \sim \mathcal{O}(10^{-1}),
\end{equation}
meaning that advective accelerations can be neglected in favor of the Coriolis effect\footnote{We note that this assumption is not satisfied across the entire parameter regime spanned by the hot Jupiter population. It is however appropriate for the hotter component of the sample where because of higher atmospheric conductivity, magnetic effects should be more appreciable.} \citep{1992phcl.book.....P}. Thus, the entire left hand side of equation (\ref{NavierStokes}) is set to zero.

Finally, we assume that the atmosphere is in hydrostatic equilibrium and that the background zonal flow is geostrophic \citep{1992aitd.book.....H}. Accordingly, to zeroth order in $\epsilon$ (i.e. neglecting the dipole tilt), equation (\ref{NavierStokes}) reads:
\begin{eqnarray}
0=- 2 \vec{\Omega} \times v_{\rm{max}}\sin(\theta)\hat{\phi} + \frac{\vec{\nabla} P}{\rho} + \vec{g}.
\label{zerothorder}
\end{eqnarray}

It remains to solve the residual terms in the Navier-Stokes equation. Given the supposed smallness of $\epsilon$, it is likely that thermal advection associated with $\ve$ does not alter the global pressure profile significantly (although this assertion depends on the thermodynamic properties of the atmosphere). Consequently, an approximate force-balance between Coriolis and Lorentz forces can be envisaged at first order in $\epsilon$. In the specified regime, to leading order in $\xi$, the individual components of of the Navier-Stokes equation take the form:
\begin{eqnarray}
0 &=& \frac{B_{\rm{p}}^2 v_{\rm{max}}}{\mu_0 \eta \rho} \cos(\theta) \sin(\phi) - \frac{\Omega}{r} \frac{\partial \Psi}{\partial \theta}  \nonumber \\
0 &=& \frac{B_{\rm{p}}^2 v_{\rm{max}}}{\mu_0 \eta \rho} \cos(\theta) \cos(\phi) - \cot(\theta)\frac{\Omega}{r} \frac{\partial \Psi}{\partial \phi}.
\end{eqnarray}
It follows from direct integration of the above expressions that they are satisfied by the stream-function:
\begin{empheq}[box=\fbox]{align}
\Psi = v_{\rm{max}} \Lambda r \sin(\theta) \sin(\phi) \hat{r}.
\end{empheq}
Evidently, the magnitude of the perturbation is controlled by the product of the field obliquity and the Elsasser number. Figure (\ref{flow}) shows the geometry of the perturbed zonal jet with $\epsilon \Lambda = 0.05$ (left panel) and $\epsilon \Lambda = 0.25$ (right panel).

\section{Discussion}

In this paper, we have considered the global circulation of a weakly ionized hot Jupiter atmosphere, subject to interaction with an inclined magnetic field. Using a perturbative approach, we identified the form of the induced current primarily responsible for the distortion of zonal winds. Subsequently, we obtained a closed-form expression for the stream-function of the non-axisymmetric flow. The above discussion demonstrates that on hot Jupiters that possess sizable, significantly tilted fields, atmospheric jets are perturbed away from purely zonal states, and the fractional deviation from axisymmetric circulation scales as $||(\epsilon \ve)/\vo|| \sim \epsilon \Lambda$.

Naturally, in order to obtain an analytic solution, it was necessary to make a series of simplifying assumptions. Specifically, our treatment relies on the dominance of the dipole component of the field over higher-order harmonics, the smallness of the field's obliquity $\epsilon$, a small Elsasser number $\Lambda$, a rotationally dominated force balance, confinement of the current to a thin-shell geometry, and the validity of a kinematic (as opposed to dynamic) treatment of the large-scale circulation. The last of these assumptions is arguably most significant, since in the contrary case of $\mathcal{R}_{\rm{M}}\gg1$, the electromagnetic skin-depth effect inherent to the atmospheric differential rotation may act to attenuate the oblique component of the field \citep{1982GApFD..21..113S,2010GeoRL..37.5201S}.

Cumulatively, this means that the derived solution is not appropriate for all choices of system parameters, and more exotic behavior is indeed possible beyond the scope of the specified limitations. Still, the derived solution may (at least qualitatively) be informative beyond its formal range of applicability\footnote{For example, an examination of the calculation performed in the Appendix suggests that the corrections which arise from retaining the $(\ve \times \Bo)$ term in the induction equation will alter the behavior of the perturbed flow to not increase without bound with $\Lambda$, but to force the obliquity of the jet to correspond to the obliquity of the field (and not exceed it). In other words, it may be speculated that instead of scaling as $\propto \Lambda$, the flow's obliquity will scale as $\propto \Lambda/(1+\Lambda)$ (or something similarly behaved) within the context of a more complicated calculation.}. Suitably, the onset of complex behavior should be investigated in detail using a numerical non-ideal MHD GCM \citep{2013ApJ...776...53B, 2014arXiv1401.5815R}.

An important quality of the performed calculation is that it is observationally significant. In the recent years, the forefront of the observational study of hot Jupiters has transitioned away from unembellished detection towards direct characterization. To this end, \citet{2007MNRAS.379..641C} and \citet{2007Natur.447..183K} performed the first observational surveys of thermal phase variations on close-in gas giants. While the focus of the former study was aimed at measuring the disparity in brightness between the dayside and the nightside on the planets HD209458b \& HD179949b (see also \citealt{2009ApJ...703..769K,2011ApJ...729...54C,2011ApJ...726...82C}), the latter study successfully derived a longitudinal thermal map of HD189733b. The interpretation of both sets of results has been instrumental in constraining the true state of hot Jupiter atmospheric dynamics (see e.g. \citealt{2009ApJ...699..564S,2010ApJ...720..344L}). 

Unlike purely zonal flows that transport heat longitudionally, magnetically-supported perturbations discussed in this work give rise to latitudinal flows as well as the associated meridional heat transport. Hence, (with the exception of specifically chosen field geometries) the dayside hot spot may be advected not only away from the sub-solar point \citep{2002A&A...385..166S} but away from the equator. Moreover, under the assumption of tidal spin-synchronization \citep{1981A&A....99..126H}, such perturbations are time-independent in a frame co-rotating with planet.

Although the aforementioned phase mapping approach \citep{2008ApJ...678L.129C} does not posses latitudinal resolution, the newly proposed eclipse mapping method \citep{2012ApJ...747L..20M} is likely to prove instrumental in constructing an aggregate of 2D exo-atmospheric profiles. To this end, the calculations of \citet{2012A&A...548A.128D} show that the localization of the planetary hot spot within the framework of conventional analysis is model-dependent. However, they have also pointed out that multi-wavelength scanning of the eclipse may yield exoplanetary portraits of enhanced resolution and the characterization of the atmospheric time-variability may be feasible. Such maps, combined with transit observations in the near-UV, that place constraints on the planetary magnetic field \citep{2010ApJ...722L.168V,2011MNRAS.411L..46V}, may directly probe the phenomena described in this work. Accordingly, observational efforts aimed at quantification of magnetically perturbed jets will be greatly aided by the commencement of next-generation space-based missions such as JWST.

\acknowledgments 
\textbf{Acknowledgments}. We thank Dave Stevenson, Nick Cowan, Tami Rogers, Adam Showman, Chris Spalding, Vivien Parmentier and Tristan Guillot for inspirational discussions. Additionally, we are thankful to the anonymous referee, whose thoughtful report led to a substantial improvement of the manuscript.

\appendix
In our computation of the the induced field, we explicitly assumed that terms involving $\ve$ contribute negligibly to the solution of equation (\ref{induction}). Having derived a functional form of the stream-function associated with $\ve$, we can now deduce the conditions under which this assumption holds. To do so, we adopt the same approach taken in section 3.2. The current density induced by the interactions between $\ve$ and $\Bo$ is:
\begin{equation}
\vec{J}_{\rm{ind}}^{(\epsilon,0)} = \frac{1}{\mu_0 \eta} \left( (\vec{\nabla} \times \Psi) \times \Bo + \vec{\nabla} \Gamma \right),
\end{equation}
where in direct analogy with $\Phi$, $\Gamma$ is the electric potential. Employing zero radial current boundary conditions at the edges of the atmospheric shell as above, the latitudinal and longitudinal dependencies of $\Gamma$ must conform to that of the radial component of $(\vec{\nabla} \times \Psi) \times \Bo$. Suitably, we adopt the following functional form for $\Gamma$: 
\begin{equation}
\Gamma = w(r) \cos(\theta) \sin(\theta) \sin(\phi).
\end{equation}

As before, the continuity equation for the current yields an ODE for $w(r)$:
\begin{equation}
5 \epsilon \frac{B_{\rm{p}}^3 R_{\rm{p}}^3 v_{\rm{max}}}{\mu_0 \eta \rho \Omega} - r^2 \big(6 w(r) - r (2w'(r) + r w''(r))\big) = 0. 
\end{equation}
With the aforementioned boundary conditions, the solution for $w(r)$ takes the form:
\begin{eqnarray}
w(r) = \frac{\epsilon B_{\rm{p}}^3 R_{\rm{p}}^3 v_{\rm{max}}  \big(3 r^5 - 2 R_{\rm{p}}(R_{\rm{p}}-\delta)(2R_{\rm{p}}-\delta)(2 R_{\rm{p}}^2 - 2 \delta R_{\rm{p}}+ \delta^2) +5r (5R_{\rm{p}}^4 - 10 \delta R_{p}^3 + 10 \delta^2 R_{p}^2 - 5 \delta^3 R_{p} + \delta^4) \big)}{4 \mu_0 \eta \rho \Omega r^3 \big(5 R_{\rm{p}}^4 - 10 \delta R_{\rm{p}}^3 + 10 \delta^2 R_{\rm{p}}^2 - 5 \delta^3 R_{\rm{p}} + \delta^4 \big)}.
\end{eqnarray}
Having obtained an expression for $\vec{J}_{\rm{ind}}^{(\epsilon,0)}$, we can now write down the associated Lorentz force, which arises from the interaction of this current with the axisymmetric component of the field:
\begin{eqnarray}
\vec{F}_{\rm{L}\ 0}^{(\epsilon,0)} = \epsilon \frac{\vec{J}_{\rm{ind}}^{(\epsilon,0)} \times \Bo}{\rho}.
\end{eqnarray}
Upon vertical averaging, to leading order in $\xi$ we obtain: 
\begin{eqnarray}
\langle \vec{F}_{\rm{L}\ 0}^{(\epsilon,0)} \rangle &=& - 2 \epsilon \left( \frac{B_{\rm{p}}^2}{\mu_0 \eta \rho \Omega} \right) \frac{B_{\rm{p}}^2 v_{\rm{max}}}{\mu_0 \eta \rho} \cos(\theta)^2 \cos(\phi) \hat{\theta} + 2 \epsilon \left( \frac{B_{\rm{p}}^2}{\mu_0 \eta \rho \Omega} \right) \frac{B_{\rm{p}}^2 v_{\rm{max}}}{\mu_0 \eta \rho} \cos(\theta) \sin(\phi) \hat{\phi}.
\end{eqnarray}
Consequently, the fractional magnitude of the effect we neglected in section 3 is of order 
\begin{eqnarray}
\frac{\langle \vec{F}_{\rm{L}\ 0}^{(\epsilon,0)} \rangle}{\langle \vec{F}_{\rm{L}\ 0}^{(0,\epsilon)} \rangle} \sim \left( \frac{B_{\rm{p}}^2}{\mu_0 \eta \rho \Omega} \right) = \Lambda.
\end{eqnarray}
Evidently, the derived solution applies when the Elsasser number is substantially smaller than unity.

\end{document}